\documentclass[12pt]{article}
\usepackage{a4wide,graphicx,amsmath,amssymb,hyperref}
\usepackage[numbers,sort&compress]{natbib}
\usepackage{hypernat}
\usepackage{subfigure}
\usepackage{multirow}
\usepackage{color}
\usepackage{ulem}

\renewcommand{\vec}[1]{\boldsymbol{#1}}

\def\beq{\begin{equation}}
\def\eeq{\end{equation}}
\def\bea{\begin{eqnarray}}
\def\eea{\end{eqnarray}}

\newcommand{\newc}{\newcommand}

\newc{\scale}{0.9}
\newc{\twoscale}{0.85}
\newc{\twoFeynScale}{0.45}

\newc{\Quote}[1]{``#1''}
\newc{\cf}{\textit{cf}.~}
\newc{\ie}{i.e.~}
\newc{\wrt}{w.r.t.~}
\newc{\eg}{e.g.~}
\newc{\etc}{{\it etc}}

\newc{\thalfGe}{T_{1/2}(^{76}{\rm Ge})}

\newc{\Ge}{^{76}{\rm Ge}}

\newc{\Anubb}{{\mathcal A}_{\nubb}}

\newc{\ifb}{\textrm{fb}^{-1}}

\newc{\ab}{{\rm ab}}
\newc{\fb}{{\rm fb}}
\newc{\pb}{{\rm pb}}
\newc{\nb}{{\rm nb}}
\newc{\mb}{{\rm mb}}
\newc{\eV}{{\rm eV}}
\newc{\MeV}{{\rm MeV}}
\newc{\GeV}{{\rm GeV}}
\newc{\TeV}{{\rm TeV}}

\newc{\muf}{\mu_{F}}
\newc{\mur}{\mu_{R}}
\newc{\rts}{\sqrt{s}}

\newc{\nubb}{0\nu\beta\beta}

\newc{\lamp}{\lambda'}
\newc{\mw}{M_W}
\newc{\mee}{m_{ee}}
\newc{\lmssm}{\lambda'{\rm MSSM}}

\newc{\ellp}{e^+}
\newc{\ellm}{e^-}
\newc{\mup}{\mu^+}
\newc{\mum}{\mu^-}
\newc{\sel}{\tilde{e}_L}
\newc{\selp}{\tilde{e}^+_L}
\newc{\selm}{\tilde{e}^-_L}
\newc{\sul}{\tilde{u}_L}
\newc{\sdr}{\tilde{d}_R}
\newc{\schi}{\tilde{\chi}^0}
\newc{\sgo}{\tilde{g}}
\newc{\wprime}{W'}

\newc{\brjj}{{\mathcal{BR}(\sel\to jj)}}
\newc{\brechi}{{\mathcal{BR}(\sel\to e\schi)}}
\newc{\gammajj}{{\Gamma(\sel\to jj)}}
\newc{\gammaechi}{{\Gamma(\sel\to e\schi)}}

\newc{\sigjjjj}{\sigma(\ellm\ellm\to 4j)}

\newc{\pt}{p_{T}}
\newc{\vpt}{\vec{\pt}}
\newc{\shat}{\hat{s}}
\newc{\rtshat}{\sqrt{\shat}}
\newc{\qsq}{\langle q^2\rangle}

\newc{\mchi}{m_{\schi}}
\newc{\mgo}{m_{\sgo}}
\newc{\msel}{m_{\sel}}
\newc{\msul}{m_{\sul}}
\newc{\msdr}{m_{\sdr}}
\newc{\msq}{m_{\tilde{q}}}

\newc{\br}{\mathcal{BR}}
\newc{\ord}{\mathcal{O}}

\newc{\madgraph}{\texttt{MADGRAPH}~}
\newc{\madgraphv}{\texttt{MADGRAPH5 v1.3.2}~}

\begin{document}
  \titlepage
  \begin{flushright}
    Cavendish-HEP-2011/22\,, DAMTP-2011-77
  \end{flushright}
  \vspace*{0.5cm}
  \begin{center}
    {\Large \bf Four--jet final state in same--sign lepton colliders
      and neutrinoless double beta decay mechanisms}\\
    \vspace*{1cm}
    \textsc{C.H.~Kom$^{a,b}$ and W.~Rodejohann$^c$}\\
    \vspace*{0.5cm}
            {\it
              $^a$Cavendish Laboratory, University of Cambridge, CB3 0HE, UK\\
              $^b$Department of Applied Mathematics and Theoretical Physics, University of Cambridge, CB3 0WA, UK\\
              $^c$Max--Planck--Institut f\"ur Kernphysik, Postfach 103980, D--69029 Heidelberg Germany
            }
  \end{center}
  \vspace*{0.5cm}
  \begin{abstract}
    If neutrinoless double beta decay is observed, it will be
    important to understand the mechanism(s) behind this process.
    Using a minimal supersymmetric extension to the Standard Model
    in association with a lepton number violating coupling as an
    example, we show that if neutrinoless double beta decay is
    mediated by new TeV scale particles, looking for four--jet final
    states in future linear colliders operating in the same--sign
    electron mode could provide important information on the
    underlying mechanisms.  We study the prospects for observing this
    process at the proposed ILC and CLIC energies, and discuss the
    complementarity between such a four--jet signal and other collider
    signatures at the LHC.
  \end{abstract}


\section{Introduction}\label{sec:intro}

The observation of neutrinoless double beta decay ($\nubb$) would be
an important step towards understanding the structure of physics beyond the Standard 
Model \cite{Avignone:2007fu,Rodejohann:2011mu,GomezCadenas:2011it}. It would prove that
lepton number, an accidental symmetry of the Standard Model, is
violated. After establishing this, 
the next task would be to identify the underlying mechanism of
$\nubb$. 

The standard interpretation of $\nubb$ is the so--called light (Majorana neutrino) mass mechanism.
In this mechanism, the $\nubb$ amplitude is proportional to
a neutrino mass term that violates lepton number explicitly by 2 units.  If this is the only
source of lepton number violation (LNV) in $\nubb$, 
the rate of $\nubb$ can further provide information on the 
neutrino mass scale and ordering realised by nature, and other 
neutrino properties. Within the standard interpretation,
$\nubb$--experiments become neutrino experiments.

However, if there are additional sources of LNV (non--standard
interpretations), they could dominate the amplitude when compared with
contributions from the light mass mechanism.  For example, $\nubb$ can
be mediated by models with heavy Majorana neutrinos \cite{heavynubb},
Higgs triplet models \cite{Mohapatra:1981pm}, left--right symmetric
extensions to the Standard Model \cite{Mohapatra:1980yp}, Leptoquarks
\cite{Hirsch:1996ye}, and supersymmetric models that violate lepton
number \cite{rpvnubb}.  In the context of nuclear physics experiments,
these possibilities are expected to be differentiated to certain
extents by measuring half--life ratios of different isotopes, angular
or energy correlations between the electrons from $\nubb$, nuclear
decay to excited states, and/or electron capture. A recent review on
the different proposed mechanisms and possibilities to distinguish
them from each other can be found in Ref.~\cite{Rodejohann:2011mu}.

Another approach, which we will follow here, is to search 
for other processes in which the underlying physics of $\nubb$ is present. 
Depending on the underlying LNV model, signatures related to $\nubb$
might also be observed at colliders with TeV scale collision energies.
This is based on the observation that the lower limit on $\nubb$ 
half--lives, for example that of $\Ge$ ($\thalfGe$) measured in the
Heidelberg--Moscow experiment \cite{KlapdorKleingrothaus:2000sn}, 
\begin{equation}\label{eq:HMlimit}
  \thalfGe \ge 1.9\cdot 10^{25}\,\,{\rm yrs} \, , 
\end{equation}
corresponds generically to an amplitude with $\ord(1\,{\rm TeV})$ scale
and $\ord(1)$ dimensionless LNV couplings. This is obvious from estimating 
the amplitude for light neutrino exchange, 
\begin{equation} \label{eq:am_SI}
{\cal A}_{\rm light\,\nu} \simeq G_F^2 \frac{m_{ee}}{\langle q^2 \rangle} \sim {\rm TeV}^{-5}\, , 
\end{equation}
where $G_F$ is the Fermi constant, $m_{ee}$ the effective mass set to its upper limit 
of order 0.5 eV, and $\langle q^2 \rangle$ the squared 
momentum transfer in the process, which is of order 0.01 GeV$^2$.

If the TeV scale corresponds to the mass of the particles involved in
$\nubb$, they could be produced at colliders such as the LHC and
future linear colliders.  If these particles are (not) observed in
certain LNV signals, the information could be used to (dis--)favour
particular TeV scale $\nubb$ models.  Furthermore, studying $\nubb$
mechanisms through related collider processes allows direct access to
the underlying physics at the particle level, separated from the notorious 
complication in the nuclear physics calculation, the latter of which
is however necessary when computing $\nubb$ decay rates.

At parton level, collider processes that are the most closely related
to $\nubb$ should violate electron number by two units, involve four
first generation (initial and/or final state) quarks and no missing
energy, for instance, due to the presence of neutrinos.  This is
because the relevant Feynman diagrams could be re--interpreted as
$\nubb$ diagrams after appropriate crossings.  A consequence is that
these related processes are controlled by the same model parameters,
\ie the same particle masses and (LNV) couplings.  As an example,
Refs.~\cite{Allanach:2009iv,Allanach:2009xx} show that the observation
of same--sign di--electron final state in association with two jets at
the 14 TeV LHC could lead to predictions of $\nubb$ half--lifes, when
interpreted in the context of a minimal supersymmetric extension to
the Standard Model (MSSM) in association with a LNV coupling
$\lamp_{111}$ ($\lmssm$).  Strategies to relate LHC observables to
$\nubb$ in left--right symmetric theories have also been discussed
recently \cite{Tello:2010am}.

In a lepton collider, the signal most directly related to $\nubb$
would be
\begin{equation}
  \ellm\ellm \to 4j\,,
\end{equation}
when all four jets are originated from first generation
(anti--)quarks. This is the collider signature we shall focus on, and
is called the \Quote{4j signal} in the rest of the paper.  Clearly,
the 4j signal has {\it no} irreducible Standard Model (SM) background,
since the SM conserves lepton number.  The observation of $\nubb$ in
one or more of the many upcoming experiments (see
Ref.~\cite{GomezCadenas:2011it} for a recent review on their status
and properties) would be a strong motivation to run a linear collider
in a like--sign mode.  In $\nubb$ mechanisms involving two SM
$W$--bosons, the 4j signal arises from the \Quote{inverse $\nubb$}
\cite{Rizzo:1982kn,Belanger:1995nh}, defined as
\begin{equation}
  \ellm\ellm \to W^-W^-\,,
\end{equation}
followed by hadronic decays of the $W$'s into first generation quarks
(see Ref.~\cite{Rodejohann:2010jh} for a recent study of this
process).  Since the branching ratio of the $W$ decaying into jets is
known, knowledge of the inverse $\nubb$ cross section allows direct
inference on the $\nubb$ decay rate.  However, in the presence of new
particle(s), their branching ratio(s) into jets must also be
determined.  This leads us to the studying of the four--jet final
states.

The difference between the inverse $\nubb$ followed by hadronic $W$
decays and the more general 4j process can be important, as the source
of LNV can be contained in different stages of the four--jet production.
For inverse $\nubb$, the source of LNV must be contained within the
generation of the same--sign $W$ pairs.  This is the case, \eg for the
light mass mechanism or the SM with the addition of heavy Majorana
neutrinos.  In the case of $\lmssm$, a pair of same--sign selectrons 
($\sel$), the superpartner of a left--handed electron, can be
produced via {\it gauge} interactions with an $t$--channel neutralino
($\schi$):
\begin{equation}
  \ellm\ellm \to \sel^-\sel^- \, .
\end{equation}
This can be followed by LNV decays of the selectrons into first
generation quarks via the $\lamp_{111}$ coupling, leading to the 4j
signal. Note that the $\Delta L = 2 $ process is mediated by two $\Delta
L = 1$ vertices, and not with an explicit $\Delta L = 2 $ mass term as
for processes involving Majorana light or heavy neutrinos.  The
presence of two intermediate {\it on--shell} particles in the 4j
signal could hence be an important clue to the underlying $\nubb$
mechanism.  The 4j signal cross section could also be enhanced, when
the 2--to--2 $\sel$ pair production process is followed by decay into
four jets with large branching ratios due to a large $\lamp_{111}$.  This
should be compared with the light and heavy mass mechanisms, for which
the stringent $\nubb$ half--life limit implies a relatively small 4j
signal rate \cite{Belanger:1995nh,Rodejohann:2010jh}.

In this paper, we investigate whether in the context of $\lmssm$ the
4j signal might be observable in future linear colliders, assuming
$\nubb$ is measured in the next generation of experiments.  We shall
study which $\schi$--$\sel$ mass regions future linear colliders might
be particularly sensitive to, and the effect of the presence of
competing $\sel$ decay channels, in particular gauge decay of $\sel$
into an electron and $\schi$.  Implications of like--sign linear
  collider for other (lepton number conserving) SUSY searches have
  been discussed in Refs.~\cite{likeSignColliderSUSY}.
We note that some results in this paper may also be relevant to 
left--right symmetric models, or other SM extensions with TeV scale $W'$
and heavy Majorana neutrinos which play the role of the $\sel$ and
$\schi$.

Since related collider signatures could also be expected at the LHC,
we also comment on how the 4j signal might complement other $\nubb$
probes at the LHC.

The paper is organised as follows: in the next section, we introduce
the $\lmssm$ model, and discuss features that the Feynman diagrams
possess which could enhance the $\ellm\ellm\to 4j$ signal when
compared with selected models, given the current $\nubb$ limit.  We then
explore quantitatively regions of parameter space where such
enhancements are significant, and how the prospect of observation
could change in light of future $\nubb$ data, before briefly
discussing the complementarity with some related LHC signals.


\section{The $\lmssm$ model}\label{sec:kinematics}

The $\lmssm$ model that we consider includes a (gauge invariant) LNV
superpotential term 
\begin{eqnarray}\label{eq:lampSuperpot}
  \mathcal{W}_{\lamp_{111}}&=&\lamp_{111}L_1Q_1D^c_1\,,
\end{eqnarray}
in addition to the standard MSSM interactions that conserve
$R$--parity.  The $L_1$, $Q_1$ and $D^c_1$ are the first generation
lepton doublet, quark doublet and down quark singlet superfields,
respectively.  This and all $R$--parity conserving interactions are
invariant under a discrete $\mathcal{Z}_3$ symmetry \cite{z3}.  Because
baryon number is not violated in $\lmssm$, the proton is stable.  The
Yukawa potential derived from the above LNV superpotential includes
interactions for the field combinations
\begin{equation}\label{eq:lampLagrangian}
  \tilde{l}_1q_1d^c_1\,,\qquad
  l_1\tilde{q}_1d^c_1\,,\qquad
  l_1q_1\tilde{d}^c_1\,,
\end{equation}
where the fields with (without) tildes are the supersymmetric (SM)
particles in self--evident notations.  These interactions violate
electron number by one unit, with interaction strengths proportional
to the coupling $\lamp_{111}$.  Also relevant for our discussion are
the neutralinos ($\schi$) and gluinos ($\tilde{g}$), which are the
superpartners of the neutral (gauge) bosons and gluons, respectively.
They interact with the SM particles with gauge interaction strengths.

\begin{figure*}[!t]
  \begin{center}
    \scalebox{0.52}{
      \includegraphics{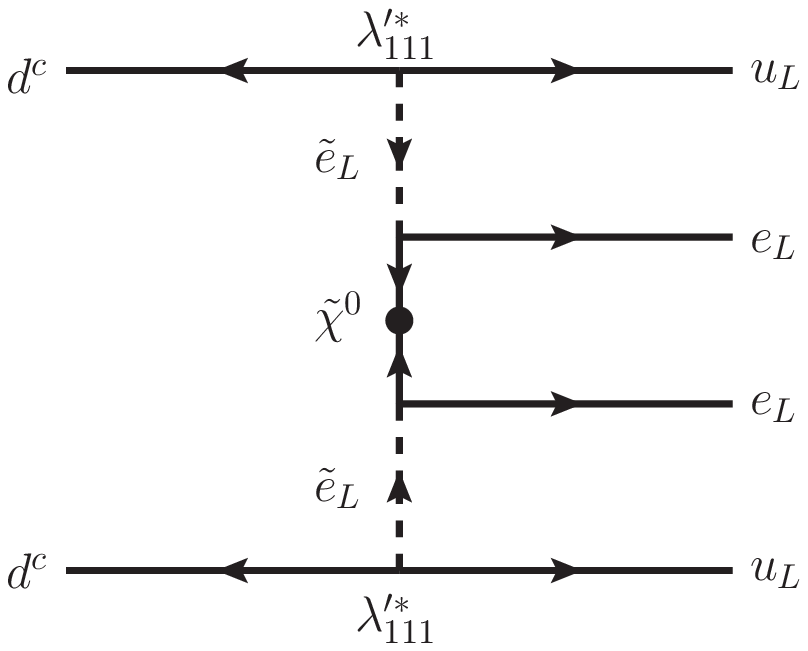} 
      \includegraphics{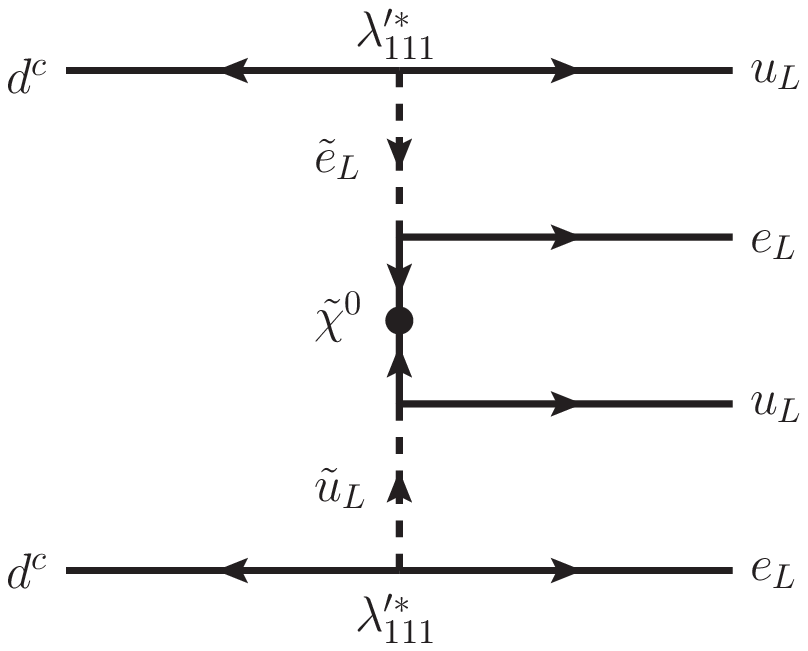} 
      \includegraphics{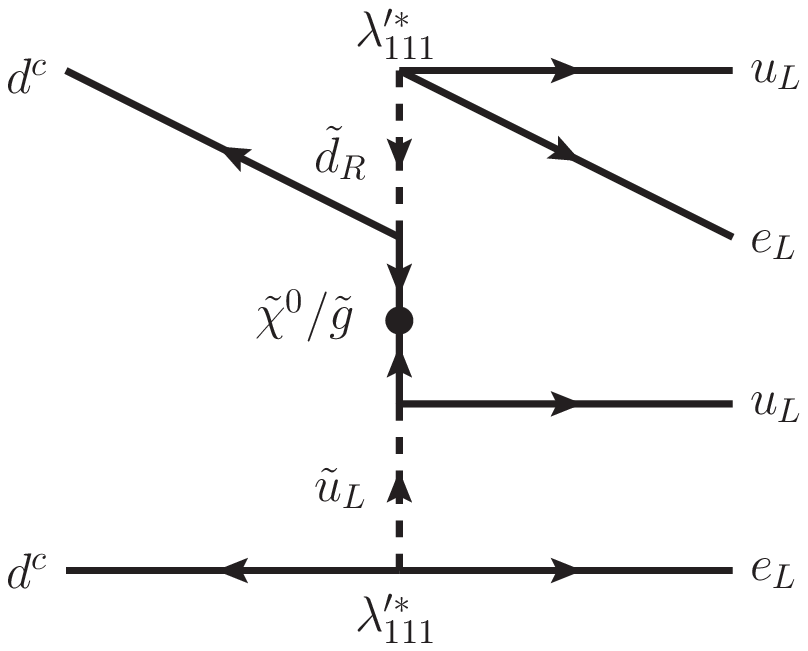} 
    }
    \caption{Example $\nubb$ Feynman diagrams for $\lmssm$.  The
      initial (final) states are on the left (right) side of the
      diagrams.  The arrows denote the flow of left chirality, and the
      dots represent mass insertions in the fermion propagators
      required by the chiral structure.  The LNV vertices are labelled
      with $\lamp_{111}$.}
    \label{fig:diag_RPV0vbb}
  \end{center}
  \begin{center}
    \scalebox{0.52}{
      \includegraphics{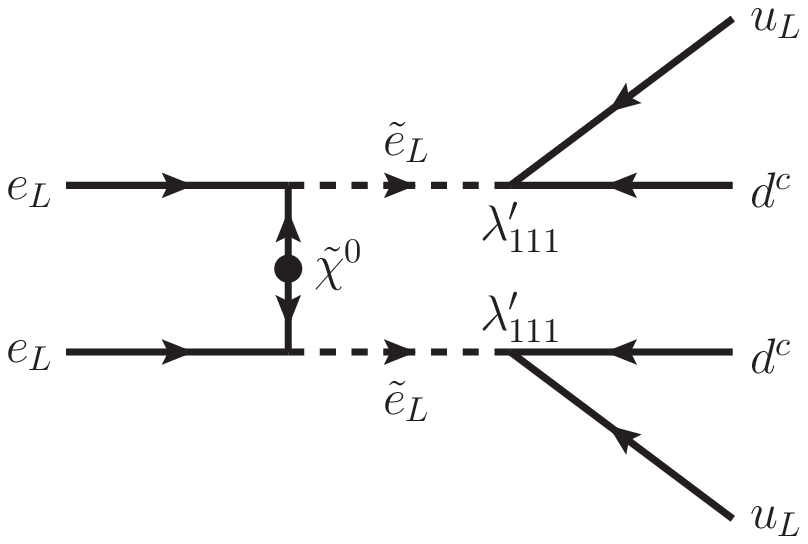} 
      \includegraphics{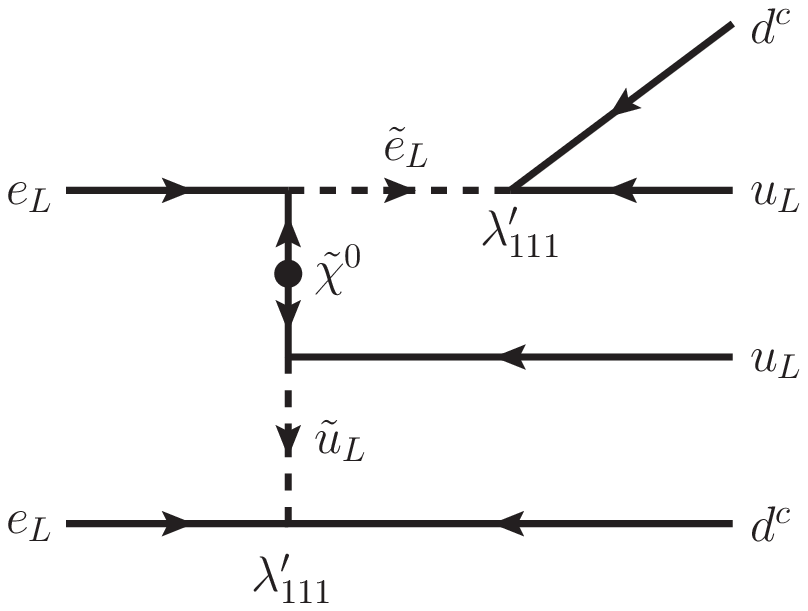} 
      \includegraphics{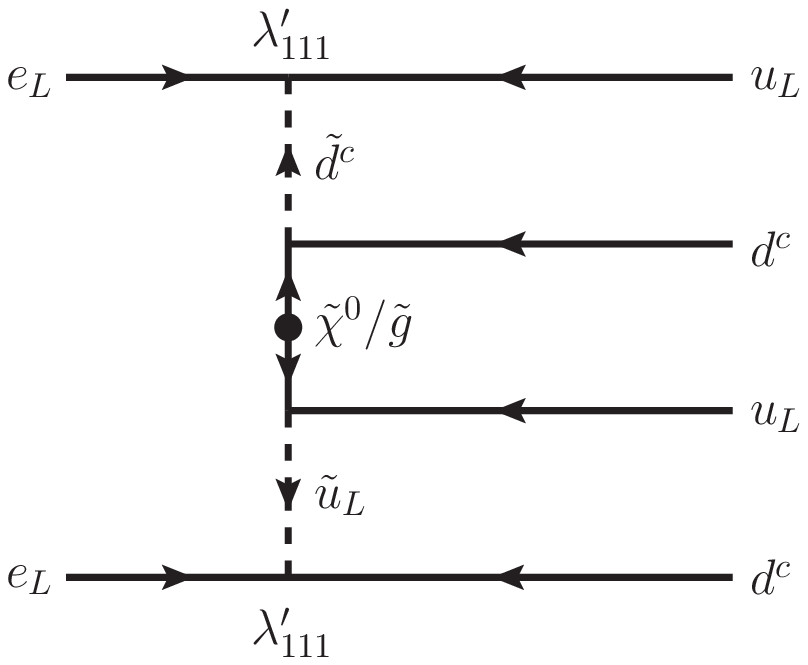} 
    }
    \caption{Example 4j signal Feynman diagrams for $\lmssm$ that
      correspond to the $\nubb$ diagrams displayed in
      Fig.~\protect{\ref{fig:diag_RPV0vbb}}. The initial (final)
      states are on the left (right) side of the diagrams.  The arrows
      denote the flow of left chirality, and the dots represent mass
      insertions required by the chiral structure.  The LNV vertices
      are labelled with $\lamp_{111}$.}
    \label{fig:diag_RPV4j}
  \end{center}
\end{figure*}

In Fig.~\ref{fig:diag_RPV0vbb}, we show example $\nubb$ Feynman
diagrams in $\lmssm$.\footnote{We note that there is another $R$--parity
violating (long--range) diagram for $\nubb$ which depends on
$\lambda'_{113} \lambda'_{131}$. This can be tested via
$B_d^0$--$\overline{B}_d^0$ mixing \cite{Allanach:2009xx}.}  
The corresponding Feynman diagrams for the 4j
signal are displayed in Fig.~\ref{fig:diag_RPV4j}. We see that the
two sets of diagrams can be related by crossing between external legs,
indicating the model parameters entering the two processes are the
same.  However, the 4j signal will be dominated by the diagram
involving two intermediate selectrons if they can be produced
{\it on--shell}, see the left diagram of Fig.~\ref{fig:diag_RPV4j}.  In this
case the partial width and branching ratio of the selectron decaying
into jets depend indirectly on the squark and gluino masses via
$\lamp_{111}$, the latter of which is extracted from an observed
$\nubb$ half--life value.  In what follows, we shall focus primarily on
the case when the selectrons are produced on--shell.

Since the momentum transfer in $\nubb$ is much smaller than the
sparticle masses involved, the effective operators for $\nubb$ are
dimension nine operators involving six fermions.  The lower $\thalfGe$
bound in Eq.~(\ref{eq:HMlimit}) leads to the approximate limit 
\cite{Hirsch:1995ek,Faessler:1998qv,Faessler:1999zg}
\begin{equation}\label{eq:lmssm}
  |\lamp_{111}| \lesssim
  5\cdot10^{-4}\left(\frac{m_{\sel,\sul,\sdr}}{100\,{\rm
      GeV}}\right)^2\left(\frac{m_{\tilde{g},\schi}}{100\,{\rm
      GeV}}\right)^{1/2} \, .
\end{equation}
Actually this limit can be easily understood by comparing the standard
amplitude in Eq.~(\ref{eq:am_SI}) with the expression 
${\cal A}_{\lmssm} \simeq |\lamp_{111}|^2 /(m_{\sel,\sul,\sdr}^4 \,
m_{\tilde{g},\schi})$. 
While this is the most stringent single coupling bound on
$\lamp_{111}$ for sparticle masses of $\ord(100\,\GeV)$, it relaxes
rapidly as the sparticle mass scale increases, and $\lamp_{111}$ can
be of $\ord(1)$ if the masses involved in the dominant $\nubb$
diagram(s) are of $\ord(1\,\TeV)$.

Other single coupling bounds on $\lamp_{111}$ come from atomic parity
violation and charged current universality in the lepton and quark
sectors \cite{Barbier:2004ez}.  These constraints\footnote{Note that
  there is a partial cancellation between $\lamp_{111}$ contributions
  to atomic parity violation through the up-- and down--squarks,
  leading to somewhat less stringent limits than other experiments.}
come from dimension six operators, all of which have limits of order
$|\lamp_{111}|\lesssim 0.02\,(m_{\sdr,\sul}/100\,\GeV)$.  Because of
the different mass dependence, whether the $\nubb$ bound is more
stringent or not depends on all sparticle masses involved, and has to
be calculated on a case--by--case basis.

Clearly, for the $\lmssm$ all intermediate propagators can contribute to
the suppression of the $\nubb$ rate.  For comparison, in the light mass
mechanism, the suppression is due to the ratio of the 
small mass insertion, $\mee \lesssim 0.5$ eV, 
to the energy scale of the process $\qsq \simeq 0.01$ GeV$^2$. 
For the heavy neutrino  mechanism, 
Majorana neutrinos (with mass $M_i^2 \gg \qsq$) which couple to the electron and
$W$ boson with strength proportional to the mixing parameter $S_{ei}$
provide the mass suppression, \ie
\begin{equation}\label{eq:heavyMaj}
{\cal A}_{\rm heavy\,\nu}
\propto \frac{S^2_{ei}}{M_i}\,, ~
|S_{ei}| \lesssim 2.5\cdot10^{-3}\left(\frac{M_i}{100\,\GeV} \right)^{1/2}\,,
\end{equation}
where in the second expression, the limit is obtained neglecting
possible large cancellations between different heavy neutrinos, which
would lead to less stringent limits.  With this caveat, we see that
the upper limit on $\lamp_{111}$ can be much less stringent than
$S_{ei}$, in particular, for masses of $\ord(1\,\TeV)$, because the
dependence on the relevant particle mass is different.

The role that $\lamp_{111}$ and $S_{ei}$ play in the 4j production is
also different.  Future $e^-e^-$ colliders will have sufficient
energies to produce same--sign $W$ pairs.  On the other hand, whether
same--sign $\sel$ pairs could be produced on--shell via electroweak
interaction depends on the mass of $\sel$.  If this is kinematically
allowed, this can be followed by LNV decay into jets.  For the heavy
mass mechanism, the same--sign $W$ production cross section is
approximately proportional to $|S^2_{ei}|^2$.  The production can then
be followed by the decay of the $W$'s into four (first generation)
jets with known branching ratios.  Using Eq.~(\ref{eq:heavyMaj}) we
see that the same--sign $W$ pair production cross section should be
smaller compared with on--shell same--sign $\sel$ production for
neutralino/heavy neutrino masses of $\ord(1\,\TeV)$.
For the light mass mechanism, the small $\mee$
value renders the 4j signal unobservable in practice
\cite{Belanger:1995nh,Rodejohann:2010jh}.

Since the $\lamp_{111}$ coupling can be of $\ord(1)$, the decay of
$\sel$ into jets could also be enhanced when compared with the
heavy/light mass mechanism.  More precisely, the branching ratio $\brjj$
depends on the $\lamp_{111}$ coupling and the presence of competing
decay channels, for example, through MSSM gauge interaction $\sel\to
e\schi$.  If the latter channel is open,\footnote{For simplicity, we assume
  all other sparticles to be much heavier than $\sel$.} \ie when $m_{\sel} >
m_{\schi}+m_{e}$, $\brjj$ can be suppressed for $m_{\sel}$ and
$m_{\schi}$ of $\ord(100\,\GeV)$, but will be
much larger for $m_{\sel/\schi}$ of $\ord(1\,\TeV)$ due to the scaling
of the $\lamp_{111}$ coupling, \cf Eq.~(\ref{eq:lmssm}).  However,
even in the $\ord(100\,\GeV)$ region, $\brjj$ can still be large in
the narrow band where $m_{\sel}-m_{\schi}\ll m_{\sel}$.  If the
selectron is the lightest supersymmetric particle, then in the
absence of other $R$--parity violating couplings, $\brjj=1$.

Whether $\mchi<\msel$ or $\mchi>\msel$, the mass and total width of
$\sel$ can be reconstructed by looking at the dijet invariant mass
distributions.  Together with the observation of the other $\sel$
decay channels, $\brjj$ and hence the value of $\lamp_{111}$ could be
estimated.  Furthermore, the mass of $\schi$ could be estimated using
the rate of the $\ellm\ellm\to\sel^-\sel^-$ process.

To sum up this section, we have seen qualitatively that for the
$\lmssm$ model, there are regions of parameter space where the LNV
process $\ellm\ellm\to 4j$ could be observed, despite the smallness of
the closely related $\nubb$ amplitude.  We have further argued that
the 4j cross section could be much larger than other possible $\nubb$
models, specifically the light and heavy mass mechanisms.  The
prospects for observing the 4j events also depend on the
center--of--mass energy, while an observation could allow inference of
the $\nubb$ contribution from the $\sel$ mediated diagrams.  In the
next section, we shall perform a more quantitative analysis with a
simplified $\lmssm$ model, and discuss the prospects for observing 4j
events at different center--of--mass energies, sparticle mass regions,
as well as with different (future) $\nubb$ limits.  Our study will be
at the cross section level, while a more detailed analysis, for
example mass reconstruction and precise determination of $\brjj$ is
beyond the scope of this exploratory study.

\section{Four--jet cross sections in $e^-e^-$ colliders}\label{sec:process}

In this section, we calculate the total cross section of the 4j
signal, $\sigjjjj$, in a simplified $\lmssm$ model using
\madgraphv\cite{madgraph}.  The model is obtained by extending the SM
to include the sparticles and vertices involved in the LNV interaction
terms derived from Eqs.~(\ref{eq:lampSuperpot}) and
(\ref{eq:lampLagrangian}), in addition to the relevant MSSM QCD and EW
interactions.  For concreteness, we assume that only one neutralino,
denoted $\schi$, contributes to both $\nubb$ and the 4j process, and
that $\schi$ is the Bino, the superpartner of the $U(1)_Y$ gauge
boson. The on--shell $\sel$ pair production cross section
$\sigma(e_Le_L\to\sel\sel)$ is given by \cite{WYKeung} 
\begin{equation}\label{eq:formula_ee}
\sigma(e^-_Le^-_L\to\sel^-\sel^-)=\frac{\pi\alpha^2|g_L|^4}{s}\frac{2\mchi^2}{s+2\mchi^2-2\msel^2}\left[L+\frac{2\lambda}{(s+2\mchi^2-2\msel^2)^2-\lambda^2}\right]\,,
\end{equation}
where
\begin{equation} \begin{array}{ccl} \displaystyle
  & \displaystyle L= & \displaystyle
{\rm ln}\frac{s+2\mchi^2-2\msel^2+\lambda}{s+2\mchi^2-2\msel^2-\lambda}\,,  \\
  & \lambda = & \lambda(s,\msel^2,\msel^2) = \sqrt{s^2-4s\msel^2}\,,
\end{array}
\end{equation}
and $g_L$ is the coupling between $e_L^-$, $\sel^-$ and $\schi$.  The
adjustable model parameters are
\begin{equation}
\mchi\,,\quad\mgo\,,\quad\msel\,,\quad\msul\,,\quad\msdr\,,\quad\lamp_{111}\,,
\end{equation}
while default \madgraph values for the SM parameters are used.
Specifying these parameters allows the partial widths $\gammajj$,
$\gammaechi$ and the corresponding branching ratios $\brjj$, $\brechi$
to be computed. The calculation of the 4j cross section uses
  all contributing diagrams, a subset of which is displayed in
  Fig.~\ref{fig:diag_RPV4j}.  Finite width effects are also included.

The value of $\lamp_{111}$ is taken assuming two scenarios: 
\begin{itemize}
\item $\thalfGe=1.9\cdot 10^{25}$ yrs, \ie $\nubb$ is \Quote{just
  around the corner}, \cf Eq.~(\ref{eq:HMlimit}); and
\item $\thalfGe=1.0\cdot 10^{27}$ yrs, \ie expected half--life limit
  in a ton--scale experiment.
\end{itemize}
For simplicity, a possible contribution from a $\mee$ term to $\nubb$,
which could have a different physical origin, is not included when
calculating the $\lamp_{111}$ upper limit.  This is a very valid
assumption in case neutrinos would obey a normal hierarchy (as
predicted in typical GUTs), with $m_{ee} = {\cal O}$(meV). The
extraction of $\lamp_{111}$ follows Ref.~\cite{Allanach:2009xx} (see
also Refs.~\cite{Hirsch:1995ek,Faessler:1998qv,Faessler:1999zg}),
using the nuclear matrix elements in
Refs.~\cite{Faessler:1998qv,Faessler:1999zg}.

\begin{figure*}[!t]
  \begin{center}
    \scalebox{\twoscale}{
      \includegraphics{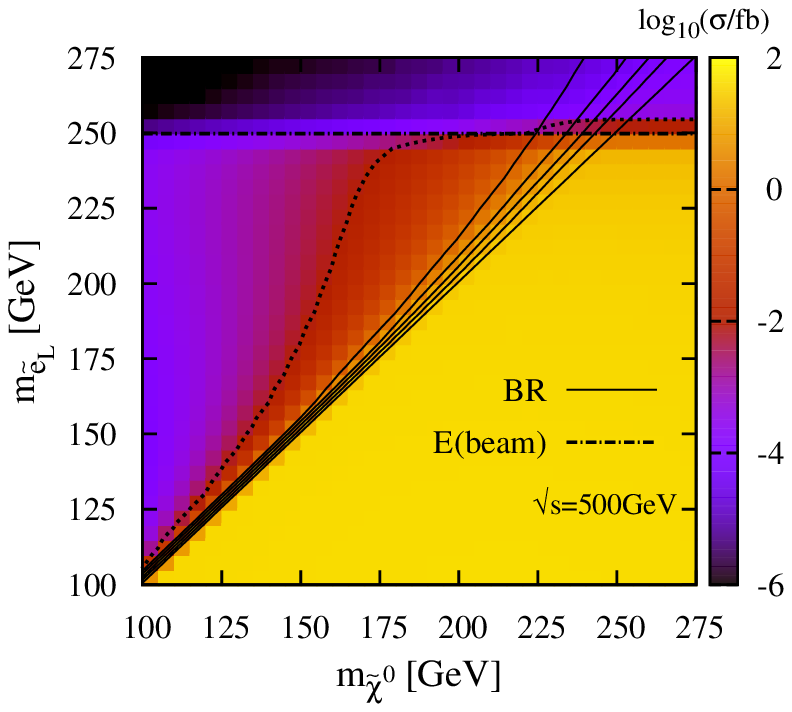}
      \includegraphics{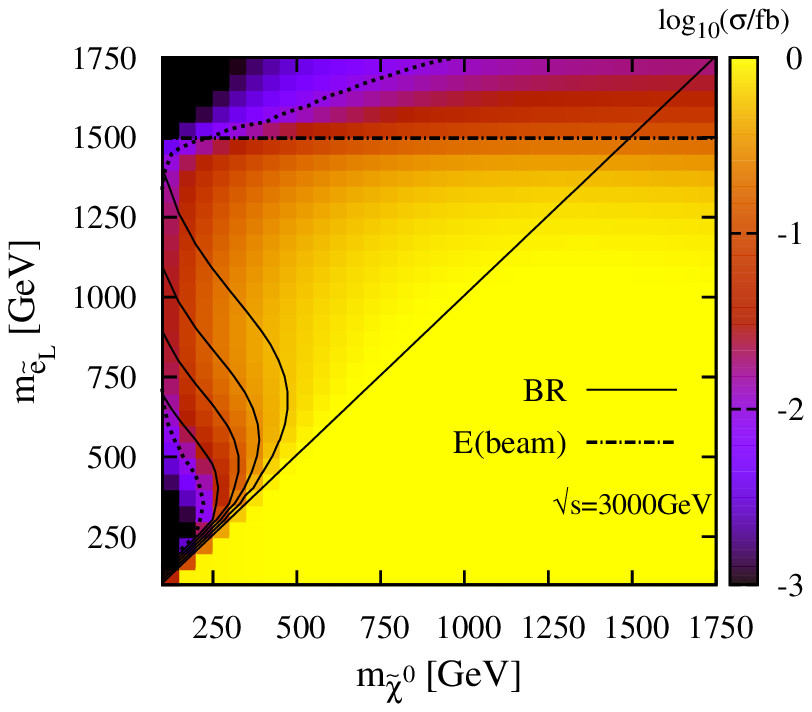} }
    \caption{Four--jet cross sections (fb) as functions of
      $(\mchi,\msel)$ at 500 GeV (left) and 3 TeV (right)
center--of--mass 
energy, assuming $\nubb$ is dominated by the $\lmssm$
      mechanism, and $\thalfGe=1.9\cdot10^{25}$ yrs.  Here the
      squark masses are set to 1000 TeV so that their contributions to
      both $\nubb$ and the 4j cross section are negligible.  In
        both panels, selectrons can be pair--produced on--shell below
        the dot--dashed line labeled E(beam).  The solid lines are branching
        fractions $\brjj$ (from right to left) from 1 to 0.2 in steps
        of 0.2.  In the region to the right of the dotted lines,
        approximately 5 or more events might be expected assuming an
        integrated luminosity of $500\,\, \ifb$.  }
    \label{fig:xsec_T1.9e25}
  \end{center}
\end{figure*}

First we consider the case for $\rts=500$ GeV, which corresponds to
the proposed ILC energy \cite{ilc}.  Using the Heidelberg--Moscow
$\nubb$ upper limit on $\lamp_{111}$, the cross section as a function
of ($\mchi$, $\msel$) is displayed in Fig.~\ref{fig:xsec_T1.9e25}
(left plot). We also show in the figure the region where
  $\sigjjjj>0.01\,\fb$.  In this region, five events or more might be
  expected assuming an integrated luminosity of
  $500\,\,\ifb$~\cite{ilc}. The squark contributions to both $\nubb$
and the 4j cross sections are decoupled by setting $(\mgo$, $\msul$,
$\msdr)=1000\,\TeV$.  Other single coupling bounds on $\lamp_{111}$
are much less stringent than the bound from $\nubb$.  In this
scenario, the mass of $\sel$ must be smaller than 250 GeV for it to be
pair--produced.  From Eq.~(\ref{eq:lmssm}), we see that this mass range
implies a much smaller value of $\lamp_{111}$ compared to the
hypercharge coupling $g_Y$.  As discussed in
section~\ref{sec:kinematics}, for $\msel>\mchi$ a large $\brjj$ is
hence possible only for a narrow band near $\msel=\mchi$.  This is
reflected in the plot by the correlation between $\sigjjjj$ and
$\brjj$, since the four jet process is well approximated using the narrow
width approximation.  In this region, $\sigjjjj$ can be of
$\ord(1-10)\,\fb$, leading to a good prospect for the observation of the
4j signal. Away from the $\msel\gtrsim\mchi$ band,
there is a significant region of parameter space where
$\sigjjjj>\ord(0.01)\,\fb$, which could also lead to a small number of
4j events.  When $\msel<\mchi$, $\brjj=1$.  In this case, the 4j cross
section is of $\ord(10-50)\,\fb$ and depends relatively mildly on the
actual $(\mchi$, $\msel)$ values, see Eq.~(\ref{eq:formula_ee}). 
The cross section falls sharply when $\msel$ increases
beyond the 250 GeV threshold, which is a simple reflection of the
narrow $\sel$ width at this mass value.
\newpage

Next we consider the higher energy option, $\rts=3$ TeV, the proposed
CLIC\footnote{The ILC and the CLIC luminosities are expected to
  be similar.  For ease of comparison the same values are used in the following.} 
energy \cite{clic}.  The cross section as a function of ($\mchi$,
$\msel$) is displayed in Fig.~\ref{fig:xsec_T1.9e25} (right plot).
Again we assume $\thalfGe=1.9\cdot10^{25}$ yrs, and set $(\mgo$,
$\msul$, $\msdr)=1000\,\TeV$.  We again delineate the region where
$\sigjjjj>0.01\,\fb$, which would lead to an expectation of five or
more events assuming an integrated luminosity of $500\,\,
\ifb$.  We
see that the linear collider can access the \Quote{natural} $\sel$ and
$\schi$ mass scale of $\ord(1)$ TeV, where the upper $\lamp_{111}$
limit from $\nubb$ is of $\ord(1)$.  Now the width $\gammajj$ can
dominate over the competitive channel $\gammaechi$ in a much larger
region of ($\msel$, $\mchi$) parameter space.  Much of the parameter
space where $\msel>\mchi$ has cross sections of $\ord(0.1)\,\fb$ or
more, leading to $\ord(50)$ or more events with an assumed luminosity
of $500\,\fb^{-1}$.  When compared with Eq.~(\ref{eq:formula_ee}),
which is applicable when $\msel<\mchi$ and when the narrow width
approximation is valid, the large $\gammajj$ leads to slightly smaller
cross sections.  The large width effect that has been included in the
calculation can also be seen in the kinematic threshold at $\msel=1.5$
TeV, where the cross section drops smoothly as the mass crosses over
this limit.

\begin{figure*}[!t]
  \begin{center}
    \scalebox{\twoscale}{
      \includegraphics{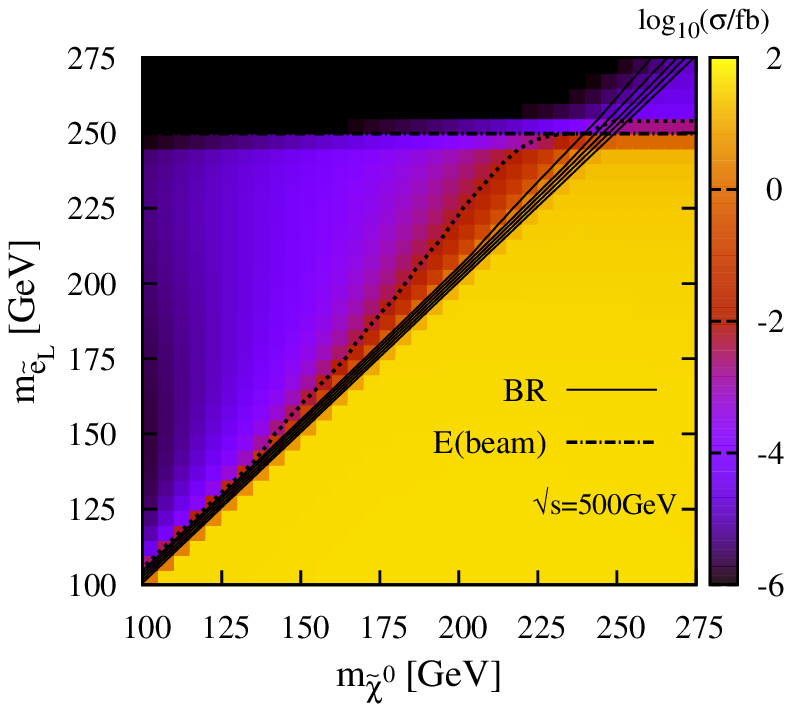}
      \includegraphics{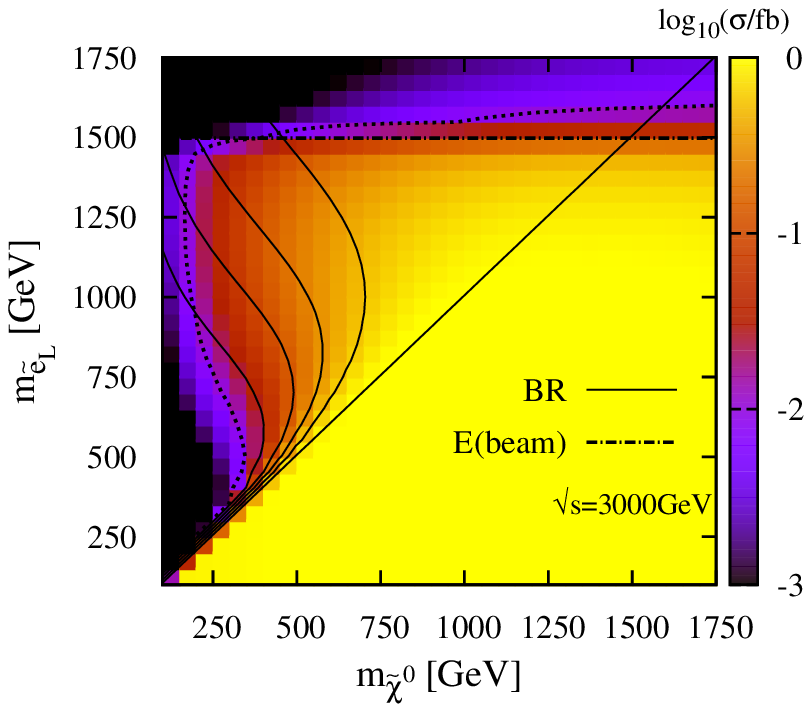}
    }
    \caption{Four--jet cross sections (fb) as functions of
      $(\mchi,\msel)$ at 500 GeV (left) and 3 TeV (right) 
center--of--mass energy, assuming $\nubb$ is dominated by the $\lmssm$
      mechanism, and $\thalfGe=1\cdot10^{27}$ yrs.  Here the squark
      masses are set to 1000 TeV so that their contributions to both
      $\nubb$ and the 4j cross section are negligible.  In both
        panels, selectrons can be pair--produced on--shell below the
        dot--dashed line labeled E(beam).  The solid lines are branching
        fractions $\brjj$ (from right to left) from 1 to 0.2 in steps
        of 0.2.  In the region to the right of the dotted lines,
        approximately 5 or more events might be expected assuming an
        integrated luminosity of $500\,\, \ifb$.  }
    \label{fig:xsec_T1e27}
  \end{center}
\end{figure*}

\begin{figure*}[!t]
  \begin{center}
    \scalebox{\twoscale}{
      \includegraphics{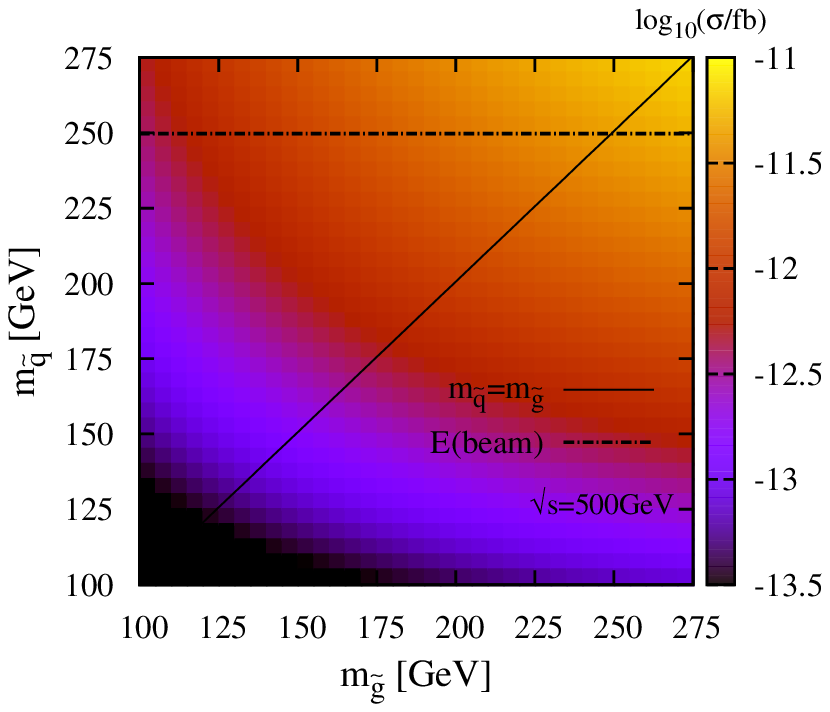}
      \includegraphics{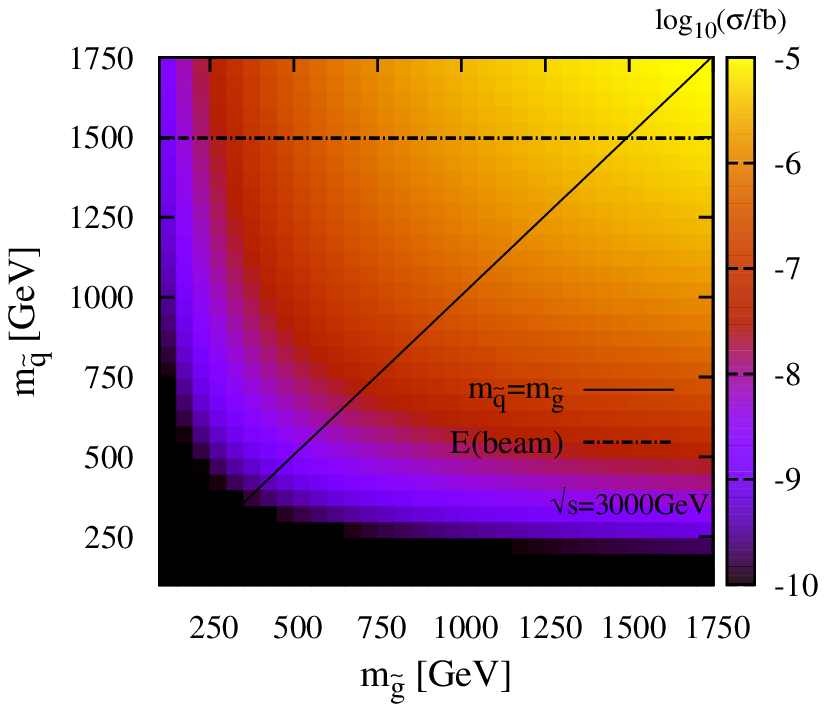}
    }
    \caption{Four--jet cross sections (fb) as functions of $(\mgo,
      \msq)$ at 500 GeV (left) and 3 TeV (right) center--of--mass
      energy, assuming $\nubb$ is dominated by the $\lmssm$ mechanism,
      and $\thalfGe=1.9\cdot10^{25}$ yrs.  Here all squark masses
      are set to a common value $\msq$.  The slepton and neutralino
      masses are set to 1000 TeV so that their contributions to both
      $\nubb$ and the 4j cross section are negligible.  The squark
      mass corresponding to $\rts/2$ and the boundary $\mgo=\msq$ are
      displayed in dot--dashed and solid lines respectively. }
    \label{fig:xsec_SQGO}
  \end{center}
\end{figure*}

We now consider how the prospect for observing the four--jet events
changes when the half--life limit increases to
$\thalfGe=1.0\cdot10^{27}$ yrs.  The cross sections for
$\rts=500\,\GeV$ and $\rts=3\,\TeV$ are shown in the left and right
plot of Fig.~\ref{fig:xsec_T1e27} respectively.  Keeping the mass
parameters constant, increasing the half--life value leads to a more
stringent upper limit on $\lamp_{111}$.  Recall that same--sign
  $\sel\sel$ pair is produced via gauge interaction with a $t$--channel
  neutralino (see the left diagram in Fig.~\ref{fig:diag_RPV4j}).
  This 2--to--2 process is hence independent of $\lamp_{111}$, so that
  this coupling only affects the partial widths and branching ratios
  of $\sel$. The change in 4j cross
section is therefore not proportional to the change in measured
$\nubb$ half--life values.  This is especially true for the region
where we expect the most sensitivity, \ie $\msel<\mchi$ and
$\msel\gtrsim\mchi$, since in these regions the total $\sel$ width is
dominated by $\gammajj$.  In fact, for $\msel<\mchi$ we have
  $\gammaechi=0$, hence $\brjj=1$ and the 4j cross section is
  independent of $\lamp_{111}$, up to finite width effects.  The value
  of $\lamp_{111}$ may be determined by measuring the $\sel$ width.

In the region $\msel>\mchi$, the 4j cross section decreases with
$\lamp_{111}$.  The impact of changing $\thalfGe$ from
$1.9\cdot10^{25}$ yrs to $1.0\cdot10^{27}$ yrs is more pronounced
for $\rts=500\,\,\GeV$, as the competing width $\gammaechi$ dominates,
and we see that the parameter regions with $\sigjjjj>0.01\,\fb$
decreases substantially in the case of $\rts=500$ GeV.  The impact is
however somewhat less pronounced for $\rts=3$ TeV, because in most of
the parameter space displayed $\brjj$ is of $\ord(1)$.  In this
context it might be argued that the 3 TeV CLIC option provides
a better opportunity to explore the $\lmssm$ $\nubb$ mechanism.

For comparison, we show in Fig.~\ref{fig:xsec_SQGO} the 4j cross
sections obtained if $\nubb$ is dominated by squark--gluino exchanges.
Here we assume instead $\msul=\msdr=\msq$, and decouple both $\nubb$
and $\sigjjjj$ contributions from $\sel$ and $\schi$.  Again, the
$\nubb$ bound is by far the most stringent in this scenario.  Since
the squarks always appear as $t$--channel propagators, the 4j cross
sections are orders of magnitude lower than the $\sel$ dominance
scenario.  The lack of intermediate particles being produced on--shell
also implies that there are no sharp changes in cross sections when
the sparticle masses cross over the \Quote{kinematic limit} of 250
(1500) GeV in the ILC (CLIC) scenario, nor over the boundary
$\msq=\mgo$.  Because of the scaling relation Eq.~(\ref{eq:lmssm}),
the cross section depends more sensitively on $\msq$ than $\mgo$.
Moreover, the cross section ratio of the slepton--neutralino dominance
and squark--gluino dominance scenarios decreases when going from
$\rts=500$ to $\rts=3000$ GeV.  In particular, we see that the cross
section is boosted by up to 6 orders of magnitude in the parameter
region that we consider, due to the quickly relaxing $\lamp_{111}$
bound from $\nubb$ when sparticle masses increase.  Note that in the
$\msq$ region around 2 TeV, the $\nubb$ bound is only marginally more
stringent than other low energy bounds discussed in section
\ref{sec:kinematics}.  The latter can be more stringent when
increasing $\msq$ beyond 2 TeV.

\section{Summary and discussions}\label{sec:conclusion}

Let us briefly comment on the complementarity between the 4j final
state and other collider searches.  At the LHC, there are signatures
that could (indirectly) constrain the $\lmssm$ $\nubb$ mechanism.  As
discussed in Refs.~\cite{Allanach:2009iv,Allanach:2009xx}, resonant
selectron production, followed by the decay chain
\begin{eqnarray}
  pp&\to&\sel \nonumber \\
  &&\sel\to e^-\schi  \\
  &&\phantom{\sel\to e^-}\schi\to e^-jj\,,\nonumber
\end{eqnarray}
leading to same--sign di--electron (SSDE) + jets final states, can be
used to test the $\lmssm$ model in the context of $\nubb$.  Note that
the resonant production process favours large $\lamp_{111}$, while its
gauge decay into an electron favours a small $\lamp_{111}$.  This is
different from the 4j signal at lepton colliders, where a large
$\lamp_{111}$ is always preferred.  Also, the SSDE signal is only
sensitive to the region $\msel>\mchi$.  At the cross section level, an
excess of SSDE events should be more easily seen when the size of
$\lamp_{111}$ balances these competing effects.  At sufficiently low
$\msel$, resonant selectron production becomes suppressed due to the
stringent $\nubb$ bound.  In addition, the softer electrons and jets
would make the search for excess SSDE events more difficult.  At
sufficiently high $\msel$, the value of $\lamp_{111}$ would become
large enough that $\brjj$ dominates over $\brechi$.  It is thus likely
that the SSDE channel is most sensitive to some \Quote{intermediate}
selectron mass region.  Additional handles such as cuts on jet
multiplicities and missing transverse energies would enhance the
signal.  The possibility of excluding regions of parameter space using
SSDE events from early LHC data is currently under investigation
\cite{SSDEproject}.

Note that the ILC is most sensitive to the relatively low mass
  region $\msel<250$ GeV and $\msel\lesssim\mchi$.  Since this region
  is dominated by the $\sel$ decay mode $\sel\to jj$, a resonant
  selectron produced at the LHC will primarily decay into two jets
\begin{equation}
\begin{array}{ccl}
  pp\to&\sel &  \\
  &\sel&\to jj\,,
\end{array}
\end{equation}
which will likely be overwhelmed by the QCD background given such low
$\msel$.  In other words, this region might be best probed by a linear
collider.  In the very high mass region, the rapidly relaxing
$\lamp_{111}$ would again imply a large dijet resonance cross section
at the LHC, while the large decay width could complicate the prospect
for (early) detection.  Early dijet resonance searches from the ATLAS
\cite{ATLAS:dijet} and CMS \cite{CMS:dijet} collaborations indicate no
significant excess in the dijet invariant mass $m_{jj}$ region $(\sim
800\,\GeV - 4\,\TeV)$, which may already be able to constrain the
parameter region allowed by the current $\nubb$ limit.  We postpone a
detailed analysis of impact of dijet exclusion on the $\lmssm$ model
in the context of $\nubb$ to a future study.  Below $m_{jj}\sim
800\,\GeV$, the QCD background might overwhelm the signal.  However,
as can be seen in Figs.~\ref{fig:xsec_T1.9e25} and
\ref{fig:xsec_T1e27}, 4j searches at CLIC would be sensitive to a
large portion of this region.  In any case, since a dijet resonance
does not violate lepton number, LNV observables such as the 4j signal
will be needed to establish the connection between $\nubb$ and the
dijet resonance, should the latter be observed in upcoming LHC data.

Last but not least, note that before running a linear collider in
same--sign electron mode, early indication of new physics from
$\lmssm$ could already appear in the $e^+e^-$ mode, as opposite--sign
$\sel^+\sel^-$ pairs can be produced via a virtual photon or MSSM
interactions.  If this results in the same 4j signature, it is
important to run in the same--sign electron mode to search for lepton
number violating effects that can be connected to $\nubb$.  The
possibility to change the polarisation of the electron beams could
further allow investigation of the left--handed nature of the 4j
process being considered.\\

To summarise, in this paper we have demonstrated the potential for
future linear lepton colliders in providing hints of underlying
$\nubb$ mechanisms by looking for four--jet events in same--sign
$e^-e^-$ collisions.  We have argued that for the $\lmssm$ model,
there are good prospects for observing these 4j events, especially
when the selectrons can be pair--produced on--shell.  However, if
$\nubb$ is dominated by squark--gluino exchange, the resulting 4j rate
would be too low to be observed.  For the former scenario, the
properties that the $\lmssm$ model possesses in order to make the 4j
cross section larger than that of other $\nubb$ mechanisms, in
particular the light and heavy mass mechanisms, was discussed.

We then went on to compute the 4j cross section for the $\lmssm$
model, assuming $\rts=500$ GeV (ILC) and $\rts=3$ TeV (CLIC)
center--of--mass energies.  We find that, if the $\sel$ and $\schi$
masses are around the 500--1500 GeV region, the 4j signal can be
observed at CLIC, while the ILC should also be sensitive to the
regions $\msel\gtrsim\mchi$ and $\msel<\mchi$ for $\msel<250\,\GeV$.
It is likely that there are regions of parameter space where the LHC
could provide indication that $\lmssm$ is relevant for $\nubb$, in
which case future linear colliders could act as further and presumably
cleaner tests of the $\lmssm$ model.  There are other regions which
can only be probed by future linear collider.  Further understanding
the potentials afforded by the LHC, and its possible synergy with
linear colliders on both $\lmssm$ and other $\nubb$ mechanisms, is an
interesting topic that should be studied in greater detail in the
future.


\section*{Acknowledgements}

We thank Tilman Plehn for a helpful discussion.  This work has been
supported in part by the Isaac Newton Trust, the STFC and a Royal
Society International Travel Grant (CHK).  WR was supported by the ERC
under the Starting Grant MANITOP and by the Deutsche
Forschungsgemeinschaft in the Transregio 27.  CHK would like to thank
the Particle and Astroparticle Physics group at MPIK Heidelberg and
the Rudolf Peierls Centre for Theoretical Physics at Oxford for
hospitality while part of the work was carried out.


\end{document}
